\documentclass[english,a4paper]{article}
\usepackage{dcolumn}
\usepackage{bm}
\usepackage{graphicx}
\usepackage{amsmath}
\usepackage{amsfonts}
\usepackage{amssymb}
\usepackage{amsthm}
\usepackage{amstext}
\usepackage{amsbsy}
\usepackage{amsopn}
\usepackage{amscd}
\usepackage{amsxtra}

\def\epsilon{\varepsilon}

\DeclareMathOperator{\ellF}{F}

\newcommand{\bec}{\begin{center}}
\newcommand{\enc}{\end{center}}
\newcommand{\be}{\begin{equation}}
\newcommand{\ee}{\end{equation}}
\newcommand{\bmi}{\begin{minipage}}
\newcommand{\emi}{\end{minipage}}

\newcommand{\bi}{\begin{itemize}}
\newcommand{\ei}{\end{itemize}}
\newcommand{\ba}{\begin{array}}
\newcommand{\ea}{\end{array}}

\everymath{\displaystyle}

\begin{document}
\title{Time-optimal selective pulses of two uncoupled spin 1/2 particles}
\author{L. Van Damme, Q. Ansel, S. J. Glaser\footnote{Department of Chemistry, Technische Universit\"at
M\"unchen, Lichtenbergstrasse 4, D-85747 Garching, Germany}, D. Sugny\footnote{Laboratoire Interdisciplinaire Carnot de
Bourgogne (ICB), UMR 6303 CNRS-Universit\'e Bourgogne-Franche Comt\'e, 9 Av. A.
Savary, BP 47 870, F-21078 Dijon Cedex, France and Institute for Advanced Study, Technische Universit\"at M\"unchen, Lichtenbergstrasse 2 a, D-85748 Garching, Germany, dominique.sugny@u-bourgogne.fr}}

\maketitle

\begin{abstract}
We investigate the time-optimal solution of the selective control of two uncoupled spin 1/2 particles. Using the Pontryagin Maximum Principle, we derive the global time-optimal pulses for two spins with different offsets. We show that the Pontryagin Hamiltonian can be written as a one-dimensional effective Hamiltonian. The optimal fields can be expressed analytically in terms of elliptic integrals. The time-optimal control problem is solved for the selective inversion and excitation processes. A bifurcation in the structure of the control fields occurs for a specific offset threshold. In particular, we show that for small offsets, the optimal solution is the concatenation of regular and singular extremals.
\end{abstract}

\section{Introduction}
Performing efficient and selective quantum state transfer by external electromagnetic fields remains a challenge of practical and fundamental interest with applications extending from atomic physics to magnetic resonance and quantum information science~\cite{glaserreview,brifreview,altafinireview,dongreview,alessandrobook}. Different analytical and numerical methods have been proposed up to date to design control fields~\cite{silver:1984,silver:1985,silver:Nature,mitschang:2005,tesiram:2005,warnking:2004,mitschang:2004,cano:2002,daemsprl,adiabaticreview,STA,STAnjp}. Among others, we can mention optimal control techniques~\cite{glaserreview} for which the selectivity problem has been addressed numerically with standard iterative algorithms~\cite{mao:1986,skinner:2012,conolly:1986,rosenfeld:1996,shi:1988,zou:2007,kobzar:2005,nakashima:2017,li:2002}. Such methods are interesting for designing efficient pulses, but their application is not completely satisfactory since there is generally no proof of the global optimality of the derived solution~\cite{grape,reichkrotov,gross}. Such a proof can be achieved by using geometric optimal control theory~\cite{alessandrobook,bonnardbook,jurdjevicbook} and the Pontryagin Maximum Principle (PMP)~\cite{pont}. In the case of a low dimensional control problem, this geometric analysis allows us to have a global view of the control landscape from which we can deduce the structure of the optimal solution and the physical limits of a given process, such as the minimum time to reach the target state. The PMP and the geometric techniques have been recently applied with success to a series of fundamental problems in quantum control, such as, to cite a few, the state to state transfer~\cite{alessandro,boscain,hegerfeldt:2013,vandamme:2014,nimbalkar:2012,sugny:2008}, the implementation of unitary gates~\cite{boozer:2012,garon,khanejaspin,khaneja:2002,yuan:2005}, the simultaneous control of different systems~\cite{romano:2015,albertini:2015,albertini:2016,assemat,vandamme:2017} and the control of two-level quantum systems or spin systems in presence of relaxation~\cite{lapertprl,contrast,mukherjee:2013,khaneja:2003,zhang:2011,sugny:2007,lapert:2010,stefanatos:2004,stefanatos:2005,stefanatos:2009,bonnard:2012}.

In magnetic resonance, a benchmark example for the selective control of spins is given by an inhomogeneous ensemble of spin 1/2 particles with different offsets~\cite{silver:1984,silver:Nature}. In this paper, we propose to investigate the simplest selectivity problem, that is the simultaneous time-optimal control of two uncoupled spins by means of magnetic fields. The two spins are assumed to be initially at the thermal equilibrium state, i.e. the north pole of the Bloch sphere. We consider in this work both the selective excitation and inversion processes for which the goal is to steer one of the two spins towards the equator or the south pole, while bringing back the other to the initial state. We derive the global time-optimal solution with a constraint on the maximum available field intensity. For a large offset difference, the optimal pulse is regular of maximum intensity. We show the existence of a bifurcation for a specific offset threshold. For smaller offset difference, the optimal solution is the concatenation of regular and singular arcs, the singular solution corresponding to zero field.

The article is organized as follows. In Sec. \ref{sec2}, we define the model system and we show how to apply the PMP in this case. Section~\ref{sec3} is dedicated to the presentation of the results. We derive the time-optimal solutions for the selective excitation and inversion of spins. We discuss how this minimum time varies as a function of the offset difference. A comparison with a direct numerical optimization is made in Sec.~\ref{sec5}. A summary of the different results obtained and prospective views are presented in Sec.~\ref{sec6}. Technical computations are reported in the appendices \ref{appA}, \ref{appB} and \ref{appC}.
\section{Time-optimal control of two uncoupled spins}\label{sec2}
\subsection{The model system}\label{sec2a}
We consider two uncoupled spin-$1/2$ particles with different offsets whose dynamics are described by the Bloch equation~\cite{levittbook,ernstbook}. If we neglect the relaxation effects then the dynamics of the spins are governed in a given rotating frame by:
\begin{equation}
\dot{\vec{M}}_i=\begin{pmatrix}
0 & \omega_i & -u_y \\ -\omega_i & 0 & u_x\\ u_y & -u_x & 0
\end{pmatrix}\vec{M}_i,\label{eqBloch}
\end{equation}
where $i=\{1,2\}$ is the index of the spins. The vector $\vec{M}_i={^t}(x_i,y_i,z_i)$ is the Bloch vector associated to the spin $i$, $u_x$ and $u_y$ are the components of the magnetic field along the $x$- and $y$- directions and $\omega_i$ the offset. By a judicious choice of the rotating frequency, we can set $\omega_1=-\omega$ and $\omega_2=+\omega$ without loss of generality. We consider that the control fields are bounded so that $u_x^2+u_y^2\leq 1$. The initial states of the dynamics are the two north poles of coordinates ${^t}(0,0,1)$.

In the time-minimum case, the PMP allows us to derive necessary conditions that the control fields must satisfy to realize the fastest state to state transfer. We introduce the pseudo-Hamiltonian $H_p=\vec{p}_1\cdot\dot{\vec{M}}_1+\vec{p}_2\cdot\dot{\vec{M}}_2$, where the $\vec{p}_i$'s are the adjoint states associated with each spin~\cite{pont,bonnardbook}. They satisfy the Hamilton's equations $\dot{\vec{p}}_i=-\partial H_p/\partial\vec{M}_i$~\cite{pont}. Substituting Eq.~\eqref{eqBloch} into the pseudo-Hamiltonian and introducing the vectors $\vec{L}_i=\vec{p}_i\times\vec{M}_i$, we can show that:
\[
H_p=u_x(L_{x_1}+L_{x_2})+u_y(L_{y_1}+L_{y_2})-\omega(L_{z_1}-L_{z_2}).
\]
The PMP states that the pulses $u_x$ and $u_y$ are optimal if they maximize $H_P$. Using the constraint $u_x^2+u_y^2\leq 1$, we deduce that the control fields are in the regular case of the form:
\begin{equation}
u_x=\tfrac{1}{r}\big(L_{x_1}+L_{x_2}\big),\quad u_y=\tfrac{1}{r}\big(L_{y_1}+L_{y_2}\big),\label{optfields}
\end{equation}
with:
\[r=\sqrt{(L_{x_1}+L_{x_2})^2+(L_{y_1}+L_{y_2})^2}.\]
Note that if $r(t)\neq 0$, we get the relation $u_x^2+u_y^2=1$, which is characteristic of regular fields. A singularity appears if $r=0$. This latter case is associated to singular control fields that we will study in Sec.~\ref{sec3}. Using Hamilton's equations, it can be shown that the angular momenta $\vec{L}_i$ satisfy a differential system of the form:
\begin{equation}
\dot{\vec{L}}_i=\begin{pmatrix}
0 & \omega_i & -\tfrac{L_{y_1}+L_{y_2}}{r} \\ -\omega_i & 0 & \tfrac{L_{x_1}+L_{x_2}}{r}\\ \tfrac{L_{y_1}+L_{y_2}}{r} & -\tfrac{L_{x_1}+L_{x_2}}{r} & 0
\end{pmatrix}\vec{L}_i.\label{eqmomcin}
\end{equation}
In the general case, any solution of this system is completely determined by the six initial conditions $\vec{L}_i(0)$. Since the control fields can be expressed in terms of  $\vec{L}_i$ through Eq.~\eqref{optfields}, the optimal pulses $u_x$ and $u_y$ are also parameterized by six parameters. The problem is then to adjust these parameters in order to realize a state to state transfer in the system~\eqref{eqBloch}. In Sec.~\ref{sec2b}, we will see that the number of parameters can be reduced to $2$, allowing us to describe the control landscape of the system and to determine the global optimum solution for any specific transfer, at least in the regular situation.
\subsection{The control landscape}\label{sec2b}
At time $t=0$, the spins are in their initial state $\vec{M}_i(0)={}^t(0,0,1)$. Since $\vec{L}_i=\vec{p}_i\times\vec{M}_i$, we obtain that $L_{z_1}(0)=L_{z_2}(0)=0$, reducing the number of parameters of Eq.~\eqref{eqmomcin} to 4. We introduce the angles $\varphi_1$ and $\varphi_2$ so that $\vec{L}_i(0)=(L_i\cos\varphi_i,L_i\sin\varphi_i,0)$ where $L_i\equiv \pm|\vec{L}_i(0)|$.
Moreover, in the regular case, we can set $\sqrt{L_1^2+L_2^2}=1$ without loss of generality by noting that a rescaling of the $\vec{L}_i$'s does not affect Eq.~\eqref{eqmomcin} and the control fields of Eq.~\eqref{optfields}. The number of parameters is therefore reduced to three. We will consider in this work some control problems in which only the relative phase of the two spins in the equatorial plane is relevant. This degree of freedom allows us to choose arbitrary the initial phase of the pulse. We assume that $u_y(0)=0$, that is $L_{y_1}(0)+L_{y_2}(0)=0$. Since $L_{y_1}(0)=L_1\sin\varphi_1$ and $L_{y_2}(0)=L_2\sin\varphi_2$, we get that $L_2/L_1=-\sin\varphi_1/\sin\varphi_2$. Finally, we obtain that the control landscape only depends on two parameters $\varphi_1$ and $\varphi_2$. The initial conditions of the system~\eqref{eqmomcin} are given in terms of these parameters by:
\begin{equation}
\begin{array}{ll}
L_{x_1}(0)=\tfrac{\cos\varphi_1}{\sqrt{1+\tfrac{\sin^2\varphi_1}{\sin^2\varphi_2}}}, & L_{x_2}(0)=-\tfrac{\tfrac{\sin\varphi_1}{\tan\varphi_2}}{\sqrt{1+\tfrac{\sin^2\varphi_1}{\sin^2\varphi_2}}},\\
L_{y_1}(0)=\tfrac{\sin\varphi_1}{\sqrt{1+\tfrac{\sin^2\varphi_1}{\sin^2\varphi_2}}}, & L_{y_2}(0)=-\tfrac{\sin\varphi_1}{\sqrt{1+\tfrac{\sin^2\varphi_1}{\sin^2\varphi_2}}},
\end{array}\label{eqlini}
\end{equation}
and $L_{z_1}(0)=L_{z_2}(0)=0$.
Each pair of angles $\{\varphi_1,\varphi_2\}$ leads to a solution of Eq.~\eqref{eqmomcin} and then to the corresponding control fields $u_x(t)$ and $u_y(t)$. This pulse is then substituted in Eq.~\eqref{eqBloch} which is integrated in order to obtain the Bloch vectors $\vec{M}_1(t)$ and $\vec{M}_2(t)$ as a function of time. In general, only a finite number of angles $\{\varphi_1,\varphi_2\}$ in $[0,2\pi]\times [0,2\pi]$ allows to realize a desired transfer. Among these pairs of values, the one which brings the system to the target state in minimum time corresponds to the global time optimal solution of the control problem.
\section{Analytical description of the optimal control fields}\label{sec3}
\subsection{The regular case}\label{sec3a}
\begin{figure}[h!]
\includegraphics[scale=0.6]{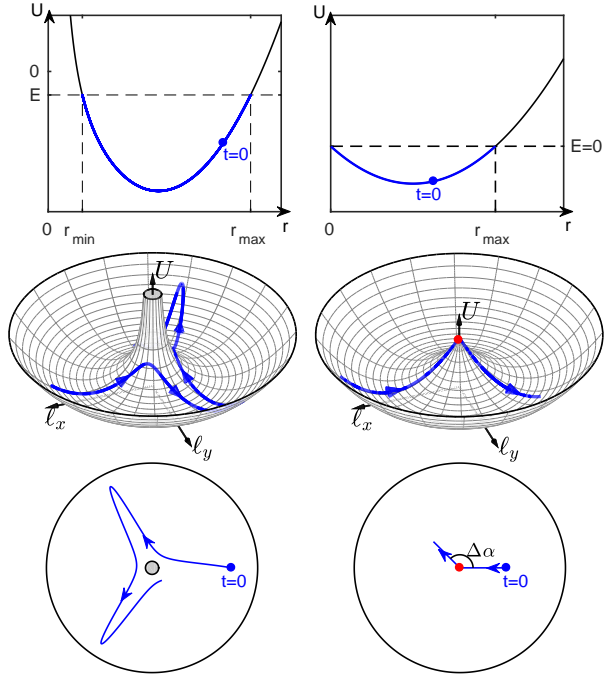}
\caption{(Color online) Trajectories of the pseudo-particle in the effective potential $U$ in the regular (left) and singular (right) cases. The upper panels show the potential $U$ as a function of the radius $r$ in black, and the motion of the particle in blue (or in dark gray).
The middle panels display the potential $U$ as a function of $\ell_x=r\cos\alpha$ and $\ell_y=r\sin\alpha$. The red dot of the right panel depicts the singular point in which the particle halts during a finite time. While the particle is stuck on this point, the control fields are singular (see Sec.~\ref{sec3b}), i.e. $u_x=u_y=0$.
The lower panels show a projection of this motion in the plane $(\ell_x,\ell_y)$, providing a better view of the variation of $\alpha(t)$, the phase of the pulse. Dimensionless units are used.\label{fig1}}
\end{figure}

We show in this paragraph that the dynamics generated by the Pontryagin Hamiltonian can be written as a one-dimensional effective mechanical Hamiltonian of a pseudo-particle moving in a potential energy landscape~\cite{arnoldbook,goldsteinbook}. This gives a geometric interpretation of the solutions of the time-optimal control problem. We focus on regular extremals in this section. We define the following coordinates:
\[\vec{\ell}=\vec{L}_1+\vec{L}_2,\quad\vec{m}=\vec{L}_1-\vec{L}_2.\]
The components of the control pulse are then $u_x=\ell_x/r$ and $u_y=\ell_y/r$ with $r=\sqrt{\ell_x^2+\ell_y^2}$.
It is straightforward to show that $\ell_z$ is a constant of motion. Moreover, since $L_{1z}(0)=L_{2z}(0)=0$ we have $\ell_z(t)=0$ for any time $t$. The dynamics of $\vec{\ell}$ and $\vec{m}$ can be written as:
\begin{equation}
\begin{cases}
\dot{\ell}_x=-\omega m_y,\\
\dot{\ell}_y=\omega m_x,\\
\dot{m}_x=-\left(\omega+\tfrac{m_z}{r}\right)\ell_y,\\
\dot{m}_y=\left(\omega+\tfrac{m_z}{r}\right)\ell_x,\\
\dot{m}_z=\tfrac{1}{r}\big(\ell_ym_x-\ell_xm_y\big),
\end{cases}r=\sqrt{\ell_x^2+\ell_y^2}.\label{eqsystlm}
\end{equation}
This system has three additional constants of motion given by:
\begin{equation}\label{eqcst}
r-\omega m_z=r_0,\quad \vec{\ell}\cdot\vec{m}=s,\quad r^2+|\vec{m}|^2=2.
\end{equation}
We set $\ell_x=r\cos\alpha$ and $\ell_y=r\sin\alpha$. The variable $\alpha$ is thus the phase of the control pulse, i.e. $u_x=\cos\alpha$ and $u_y=\sin\alpha$.
Using Eq.~\eqref{eqcst} and the relation $\alpha=\arctan(\ell_y/\ell_x)$, we arrive at:
\begin{equation}
r^2\dot{\alpha}=\omega s,\label{eqkepler}
\end{equation}
which can be interpreted as the law of equal areas in the two-body problem with a central force. The same relation holds for the Kepler's second law of planetary motion~\cite{goldsteinbook}. We deduce that the vector $\vec{\ell}$ sweeps out equal areas in equal times. Moreover, the dynamics of the system can be viewed as the motion of a pseudo-particle in a potential well. Indeed, if we define the kinetic energy term as $\dot{r}^2/2$ then it can be shown that:
\begin{equation}\label{eqnrj}
E=\tfrac{1}{2}\dot{r}^2+U(r),
\end{equation}
where the potential energy $U$ and the mechanical energy $E$ are given by:
\begin{equation}
\begin{cases}
U(r)=\tfrac{1}{2}(1+\omega^2)r^2-r_0r+\tfrac{\omega^2s^2}{2r^2},\\
E=\omega^2-\tfrac{r_0^2}{2}.
\end{cases}\label{eqpotentiel}
\end{equation}
Thus any regular time-optimal solution of the control of two uncoupled spins is associated to a trajectory of a particle of energy $E$ in a potential $U(r)$. However, note that this mechanical analogy does not directly give information about the dynamics of the original Bloch vectors $\vec{M}_i$ governed by Eq.~\eqref{eqBloch}, but only about the optimal control fields through Eq.~\eqref{eqkepler}. The dynamics of the Bloch vectors will be detailed in Sec.~\ref{sec3c}.
Using Eq.~\eqref{eqnrj}, the general solution $r(t)$ is given by the integral equation:
\begin{equation}
t=\int_{r_0}^r\tfrac{dr'}{\sqrt{2E-2U(r')}}.\label{eqr2t}
\end{equation}
The right-hand side of Eq.~\eqref{eqr2t} can be expressed in terms of a linear combination of elliptic integrals of first and third kinds together with simple analytical functions as shown in Appendix~\ref{appA1}. Only a qualitative description is given in this section. For a fixed value of $\omega$, the potential $U$ and the energy $E$ depend on the two constants of motion $s$ and $r_0$, which can be connected to the two parameters $\varphi_1$ and $\varphi_2$ defined in Sec.~\ref{sec2} through the relations:
\begin{equation}
\begin{cases}
s=\tfrac{\sin(\varphi_2-\varphi_1)\sin(\varphi_2+\varphi_1)}{1-\cos(\varphi_2-\varphi_1)\cos(\varphi_2+\varphi_1)},\\
r_0^2=\tfrac{\sin^2(\varphi_2-\varphi_1)}{1-\cos(\varphi_2-\varphi_1)\cos(\varphi_2+\varphi_1)}.
\end{cases}\label{eqsr0}
\end{equation}
Note that the parameters $s$ and $r_0$ are well-defined for any value of the angles $\varphi_1$ and $\varphi_2$.
The left panels of Fig.~\ref{fig1} show the dynamics of a pseudo-particle of energy $E$ in the potential $U$.

\subsection{Regular-singular arcs\label{sec3b}}
\begin{figure}[h!]
\includegraphics[scale=0.6]{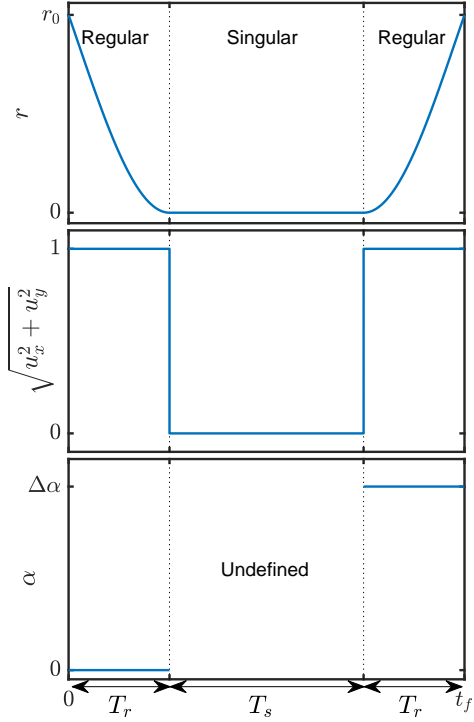}
\caption{(Color online) From top to bottom: Plot of the radius $r$, the amplitude of the control pulse $\sqrt{u_x^2+u_y^2}$ and its phase $\alpha$ as a function of time for a regular-singular solution. The two regular components have the same duration $T_r=\tfrac{\arccos(-\omega^2)}{\sqrt{1+\omega^2}}$ and the singular arc lasts during the time  $T_s$. Dimensionless units are used.\label{fig2}}
\end{figure}

According to Eq.~\eqref{optfields}, we recall that the singular fields occur when $r=0$ over a finite time duration~\cite{bonnardbook,lapertprl}. However, the effective potential of Eq.~\eqref{eqpotentiel} has a repulsive contribution given by $\tfrac{\omega^2s^2}{r^2}$ which prevents $r$ from reaching $0$. Thus, the singular fields are likely to occur only if $s=0$. As shown in Fig.~\ref{fig1}, $s=0$ is associated to a potential with a particular shape having a non-differentiable point at $r=0$ ($\ell_x=\ell_y=0$). Note also that if $s=0$, Eq.~\eqref{eqkepler} implies that $\alpha$ is constant for $r\neq 0$. The case $s=0$ involves three different types of behavior (see Fig.~\ref{fig1}):
\begin{itemize}
\item If $E<0$ ($r_0>\omega\sqrt{2}$), the particle oscillates in the potential without reaching the singular point, leading to a constant control phase $\alpha$ for any time $t$.
\item If $E>0$ ($r_0<\omega\sqrt{2}$), the particle crosses the singular point but does not stop on it. Its direction is not modified. As a consequence, $\alpha$ jumps of $\Delta\alpha=\pi$ when the particle crosses the singularity.
\item If $E=0$ ($r_0=\omega\sqrt{2}$), the singular arc appears. In this situation, the particle reaches the point $r=0$ with a zero velocity and then halts on it over a finite duration. While the particle is stuck on this point, the control fields are singular.
\end{itemize}
We focus now on the case $E=0$.
As shown in App.~\ref{appB}, the singular control fields are such that:
\[u_x^{s}(t)=u_y^{s}(t)=0.\]
Note that since $r_0$ is bounded by $\sqrt{2}$ [Eq.~\eqref{eqsr0}], the singularity cannot occur for $\omega>1$.
The exact time from which the control starts to be singular can be computed analytically. Indeed, Eq.~\eqref{eqr2t} can be easily integrated for $s=0$. When $r_0=\omega\sqrt{2}$, we have (see App.~\ref{appA3}):
\[r(t)=\tfrac{\omega\sqrt{2}}{1+\omega^2}\big(1-\cos(u)\big),\]
with $u=\sqrt{1+\omega^2}t-\arccos(-\omega^2)$. Thus, $r(t)$ is zero and singular for $t\geq t_S$ with $t_S=\tfrac{\arccos(-\omega^2)}{\sqrt{1+\omega^2}}$. A remaining fundamental question about the structure of the control protocol concerns the transition from singular to regular arcs and the number of singular arcs of the optimal trajectory. In other words, we have to compute at which time the system can exit from the singularity, and what is the corresponding phase variation. In this paper, we conjecture that the global time-optimal solutions are only composed of one singular arc. This assumption is corroborated by numerical computations as shown in Sec.~\ref{sec5}. In summary, the optimal control field is the concatenation of a regular control of phase $\alpha=0$, a zero-amplitude pulse and another regular part with a phase $\Delta\alpha$. These pulses are time-symmetric, i.e. the two regular components are of equal duration. We deduce that they depend on two parameters, namely the duration of the singular arc, denoted $T_s$ and the variation $\Delta\alpha$ of the phase. The regular-singular fields are represented in Fig.~\ref{fig2}.
\subsection{Dynamics of the Bloch vector\label{sec3c}}
In the regular case, since $\vec{L}_i=\vec{p}_i\times\vec{M}_i$, we deduce that the Bloch vector $\vec{M}_i$ is orthogonal to $\vec{L}_i$ and rotates about it with the angle $\psi_i$ whose time evolution can be computed. The momenta $\vec{L}_i$ being related to $\vec{m}$ and $\vec{\ell}$, the Bloch vectors can be expressed as a function of the coordinates of Eq.~\eqref{eqsystlm}.
As shown in Appendix~\ref{appC1}, the Bloch vectors can be written as follows:
\begin{subequations}\label{blochreg}
\begin{equation}
\begin{cases}
x_1=-\tfrac{m_z(\ell_x+m_x)\sin\psi_1}{\sqrt{2(1+s)}\sqrt{2(1+s)-m_z^2}}-\tfrac{(\ell_y+m_y)\cos\psi_1}{\sqrt{2(1+s)-m_z^2}},
\\
y_1=-\tfrac{m_z(\ell_y+m_y)\sin\psi_1}{\sqrt{2(1+s)}\sqrt{2(1+s)-m_z^2}}+\tfrac{(\ell_x+m_x)\cos\psi_1}{\sqrt{2(1+s)-m_z^2}},\\
z_1=\textstyle\sqrt{1-\tfrac{m_z^2}{2(1+s)}}\sin\psi_1,
\end{cases}
\end{equation}
\begin{equation}
\begin{cases}
x_2=\tfrac{m_z(\ell_x-m_x)\sin\psi_2}{\sqrt{2(1-s)}\sqrt{2(1-s)-m_z^2}}-\tfrac{(\ell_y-m_y)\cos\psi_2}{\sqrt{2(1-s)-m_z^2}},
\\
y_2=\tfrac{m_z(\ell_y-m_y)\sin\psi_2}{\sqrt{2(1-s)}\sqrt{2(1-s)-m_z^2}}+\tfrac{(\ell_x-m_x)\cos\psi_2}{\sqrt{2(1-s)-m_z^2}},\\
z_2=\textstyle\sqrt{1-\tfrac{m_z^2}{2(1-s)}}\sin\psi_2,
\end{cases}
\end{equation}
\end{subequations}
where the angles $\psi_i$ are solutions of:
\begin{subequations}\label{eqdpsidtreg}
\begin{equation}
\dot{\psi}_1=-\tfrac{(r^2+s)\omega^2\sqrt{2(1+s)}}{r\left[2(1+s)\omega^2-(r-r_0)^2\right]},
\end{equation}
\begin{equation}
\dot{\psi}_2=-\tfrac{(r^2-s)\omega^2\sqrt{2(1-s)}}{r\left[2(1-s)\omega^2-(r-r_0)^2\right]},
\end{equation}
\end{subequations}
with the initial conditions $\psi_1(0)=\psi_2(0)=\pi/2$. It is generally not possible to compute analytically $\psi_1$ and $\psi_2$ due to the complexity of the radius $r(t)$ which can be expressed in terms of the inverse of elliptic integrals [Eq.~\eqref{eqr2t}].


In the regular-singular case, the dynamics of the Bloch vectors are simple. The two vectors $\vec{M}_1$ and $\vec{M}_2$ are transferred from the north pole to the equator of the sphere between $t=0$ and $t=\tfrac{\arccos(-\omega^2)}{\sqrt{1+\omega^2}}$ with the first regular arc. Then, they move along the equator during the time $T_s$, and finally go to the target states driven by the second regular arc until the time $t_f$. We can show that the components of $\vec{M}_1(t_f)$ and $\vec{M}_2(t_f)$ are given by (see App.~\ref{appC2}):
\begin{subequations}\label{blochsing}
\begin{align}
\small
&\begin{cases}
\begin{aligned} x_1(t_f)=\tfrac{1}{4\omega}\big[&\cos(2\Delta\alpha-\omega T_s-\gamma)+\cos(\omega T_s+2\gamma) \\
&-\cos(2\Delta\alpha-\omega T_s)-\cos(\omega T_s+\gamma)\big]
\end{aligned}
\\
\begin{aligned} y_1(t_f)=\tfrac{1}{4\omega}\big[&\sin(2\Delta\alpha-\omega T_s-\gamma)+\sin(\omega T_s+2\gamma) \\
&-\sin(2\Delta\alpha-\omega T_s)-\sin(\omega T_s+\gamma)\big]
\end{aligned}\\
z_1(t_f)=-\cos(2\Delta\alpha-\omega T_s-\gamma)
\end{cases} \\
&\begin{cases}
\begin{aligned} x_2(t_f)=\tfrac{1}{4\omega}\big[&-\cos(2\Delta\alpha+\omega T_s+\gamma)-\cos(\omega T_s+2\gamma) \\
&+\cos(2\Delta\alpha+\omega T_s)+\cos(\omega T_s+\gamma)\big]
\end{aligned}
\\
\begin{aligned} y_2(t_f)=\tfrac{1}{4\omega}\big[&-\sin(2\Delta\alpha-\omega T_s-\gamma)+\sin(\omega T_s+2\gamma) \\
&+\sin(2\Delta\alpha+\omega T_s)-\sin(\omega T_s+\gamma)\big]
\end{aligned}\\
z_2(t_f)=-\cos(2\Delta\alpha+\omega T_s+\gamma),
\end{cases}
\end{align}
\end{subequations}
with
$\gamma=\arctan\left(\tfrac{2\omega\sqrt{1-\omega^2}}{1-2\omega^2}\right).$
Note that $\gamma$ is well defined since the singularity can play a role only if $\omega\leq 1$ (See Sec.~\ref{sec3b}).
\section{Application to the selective control of two uncoupled spins\label{sec4}}
This section presents some results for the time-optimal selective excitation (Sec.~\ref{sec4a}) and selective inversion (Sec.~\ref{sec4b}) processes. In each case, the global optimum is associated to regular control fields if $\omega$ is larger than a certain threshold $\omega_l=\tfrac{1}{2}\sqrt{2-\sqrt{2}}\simeq 0.38$ for the selective excitation and $\omega_l=1/\sqrt{2}$ for the selective inversion. When $\omega$ is smaller than this value, the optimal control field is the concatenation of regular and singular arcs as shown in Fig~\ref{fig2}. In this latter case, we have no proof of the optimality of the solution. However, a smooth transition occurs for $\omega=\omega_l$ between the two control protocols. The numerical analysis of Sec.~\ref{sec5} also strongly suggests that these solutions are the optimal ones. The regular solutions are obtained by integrating numerically the systems~\eqref{eqBloch} and \eqref{eqmomcin}.
\subsection{Selective excitation}\label{sec4a}
We consider the problem of transferring in minimum time the spin $1$ to the equatorial plane of the Bloch sphere, while bringing back the spin $2$ to its initial state at $t=t_f$. More precisely, the control problem can be defined as:
\[
\begin{aligned}
&\vec{M}_1(0)=\begin{pmatrix}
0\\0\\1
\end{pmatrix}\rightarrow \vec{M}_{1_T}=\begin{pmatrix}
x_f\\ y_f\\0
\end{pmatrix}\\
&\vec{M}_2(0)=\begin{pmatrix}
0\\0\\1
\end{pmatrix}\rightarrow \vec{M}_{2_T}=\begin{pmatrix}
0\\0\\1
\end{pmatrix},
\end{aligned}
\]
where $x_f$ and $y_f$ are free. We introduce the figure of merit $J(t)$ at time $t$:
\[
J(t)=z_1^2+(1-z_2)^2.
\]
We denote by $t_f$ the time for which $J$ is minimum. We have $J(t_f)=0$ when the transfer is exactly realized.

\begin{figure}[h!]
\centering
\includegraphics[scale=0.68]{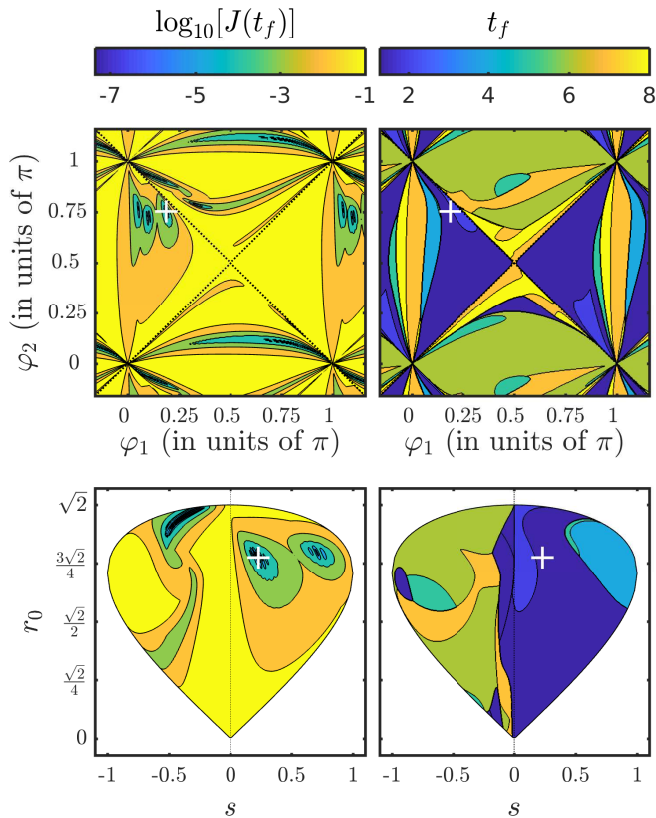}
\caption{(Color online) Figure of merit $J(t_f)$ (left) and final time $t_f$ (right) of the selective excitation process as a function of $\varphi_1$ and $\varphi_2$ (up) and $r_0$ and $s$ (down), for $\omega=1$. The dark blue (or black) regions are associated with different optima. The white cross depicts the global optimal set of parameters $\{\varphi_1^*,\varphi_2^*\}$ and $(r_0^*,s^*)$. The black dotted lines are the lines of equations $\varphi_1=\pm\varphi_2\mod\pi$ and $s=0$. Dimensionless units are used.\label{fig3}}
\end{figure}

\begin{figure}[h!]
\centering
\includegraphics[scale=0.55]{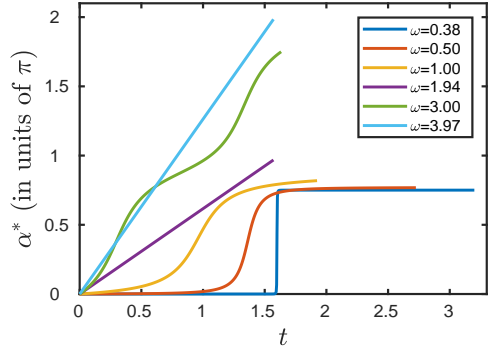}
\caption{(Color online) Time evolution of the phase $\alpha^*$ of the time-optimal control pulse for different offsets $\omega$ in the regular case of the selective excitation process. Dimensionless units are used.\label{fig4}}
\end{figure}

\begin{figure}[h!]
\centering
\includegraphics[scale=0.61]{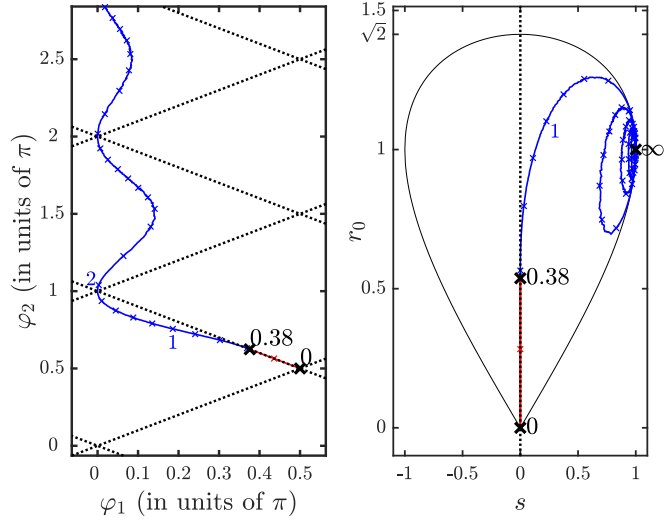}
\caption{(Color online) Evolution for the selective excitation process of the position of the global optimum $\{\varphi_1^*,\varphi_2^*\}$ (left) and $(r_0^*,s^*)$ (right) for $\omega\in [0,5]$. The regular solutions are plotted in blue (or dark gray) and the singular ones in red (or light gray).
The offset is increased by $0.2$ between each cross. The dashed lines of the left panel represent the lines $\varphi_1=\pm\varphi_2\mod\pi$, associated to $s=0$. Dimensionless units are used.\label{fig5}}
\end{figure}

\begin{figure}[h!]
\centering
\includegraphics[scale=0.5]{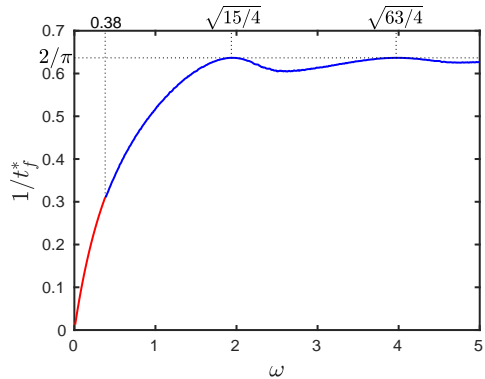}
\caption{(Color online) Plot of the inverse of the optimal time $t^*$ as a function of $\omega$. The minima ($t_f=\pi/2$) are given for some specific offsets $\omega=\tfrac{1}{2}\sqrt{16n^2-1}$ with $n\in\mathbb{N}$. The regular solutions are plotted in blue (or dark gray) and the singular ones in red (or light gray). Dimensionless units are used.\label{fig6}}
\end{figure}

\paragraph*{Regular case ($\omega>0.38$).}
To solve the regular control problem, we proceed as follow:
\begin{itemize}
\item Choose a value $\omega$ and a maximum control duration $T$. Initialize the angular momenta $\vec{L}_i(0)$ according to Eq.~\eqref{eqlini}.
\item Integrate the system~\eqref{eqmomcin} from $t=0$ to $t=T$ and compute $u_x$ and $u_y$ using Eq.~\eqref{optfields}.
\item Substitute the control fields $u_x$ and $u_y$ in Eq.~\eqref{eqBloch} and integrate it from $0$ to $T$.
\item Compute $J(t)$ for $t\in[0,T]$ and determine $t_f$ for which $J(t)$ is minimum.
\item Repeat the operation for every couple of values $\{\varphi_1,\varphi_2\}$.
\end{itemize}
Since the figure of merit depends on the control fields which depend themselves on the two parameters $\varphi_1$ and $\varphi_2$, we can visualize all the possible optima in the $(\varphi_1,\varphi_2)$- plane. Figure~\ref{fig3} shows $J(t_f)$ and $t_f$ in this plane for $\omega=1$ ($\omega_1=-1$ and $\omega_2=+1$). Note that the control landscape is $\pi$- periodic. Using Eq.~\eqref{eqsr0}, we also plot the figure of merit in the plane~$\{r_0,s\}$.
Among all the local optimal solutions, the global one is depicted by a cross in Fig.~\ref{fig3}. The optimal set of parameters is $\{\varphi_1^*,\varphi_2^*\}=\{0.1886\pi,0.7548\pi\}$ and the minimum time is $t_f^*=0.6155\pi$. This duration can be compared to the minimum time needed for a standard excitation which is equal to $\pi/2$ in the units of this paper. We plot in Fig.~\ref{fig4} the phase of the optimal pulse $\alpha^*(t)$ associated to $\{\varphi_1^*,\varphi_2^*\}$.
The same approach can be used for any other offset value. Figure~\ref{fig5} shows the position of the global optima $\{\varphi_1^*,\varphi_2^*\}$ and $\{r_0^*,s^*\}$ for any value of $\omega$.
Figure~\ref{fig6} displays the inverse of the optimal time of the process $1/t_f$ as a function of $\omega$.
Note that $t_f=\pi/2$ for some specific values of $\omega$ given by $\omega=\tfrac{1}{2}\sqrt{16n^2-1}$ with $n\in\mathbb{N}$. These values are associated to a sinusoidal pulse of the form $u_x=\cos(\omega t)$ and $u_y=\sin(\omega t)$ in resonance with the spin $1$. A time $\pi/2$ is needed with this pulse to bring the spin 1 to the equator. The spin $2$ exactly goes back to the north pole when the spin $1$ reaches the equator. Since $\pi/2$ is the minimum time for the spin 1, this time is also the minimum time for the control of the two spins. These particular pulses correspond to the parameters $s=r_0=1$ and are standard solutions in NMR (see e.g. Ref.~\cite{wider}).

In the singular case ($\omega<0.38$), we conjecture that the control pulse is regular-singular-regular as shown in Fig.~\ref{fig2}. The optimal pulse and the dynamics of the Bloch vector are the ones described in Sec.~\ref{sec3b} and~\ref{sec3c}. We point out that we do not have a rigorous proof of this statement. We have tested more complicated control structures with several singular arcs without improving the minimum time to reach the target. Numerical optimization procedure as discussed in Sec.~\ref{sec5} gives the same optimal solution. The time-symmetry of the control strategy can be understood from the dynamics of the spin 2. The first regular arc transfers the spins to the equator of the Bloch sphere. The two spins remain on the equator during the singular arc. At the end of the singular arc, the spin $2$ must come back to the north pole of the sphere. The two regular components having the same amplitude $1$, they must be of equal duration for bringing the spin $2$ to its initial position.

\begin{figure}[h!]
\centering
\includegraphics[scale=0.61]{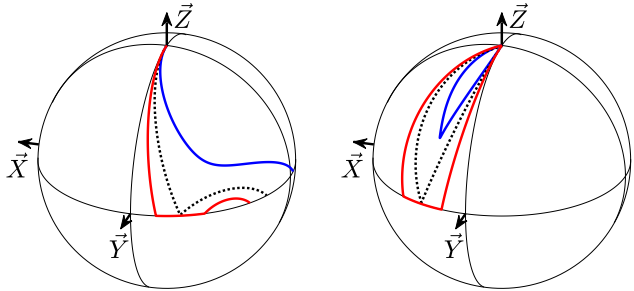}
\caption{(Color online) Time-optimal trajectories of the Bloch vectors for the selective excitation process. The left and the right panels represent respectively the spins 1 and 2. The solid-blue (or solid dark gray), dotted-black and solid-red (or solid light gray) lines correspond to $\omega=0.7$ (regular case), $\omega=\omega_S=0.38$ (limit case) and $\omega=0.2$ (singular case), respectively\label{fig7}. Dimensionless units are used.}
\end{figure}

The selective excitation is obtained by solving $z_1(t_f)=0$ and $z_2(t_f)=1$ in Eq.~\eqref{blochsing}. We get:
\begin{equation}
\begin{cases}
\Delta\alpha-\omega T_s-\gamma=\pm\frac{\pi}{2},\\
\Delta\alpha+\omega T_s+\gamma=\pm\pi.
\end{cases}
\end{equation}
Note that the + sign has to be chosen in order to have a time-minimum process. The pulse is then characterized by:
\begin{equation}
\begin{cases}
\Delta\alpha=\tfrac{3\pi}{4},\\
T_s=\tfrac{\pi}{4\omega}-\tfrac{1}{\omega}\arctan\left(\tfrac{2\omega\sqrt{1-\omega^2}}{1-2\omega^2}\right),\\
t_f=\tfrac{2\arccos(-\omega^2)}{\sqrt{1+\omega^2}}+T_s.
\end{cases}\label{solsingexc}
\end{equation}
It is straightforward to show that the duration $T_s$ of the singular arc is zero if $\omega=\omega_S=\tfrac{1}{2}\sqrt{2-\sqrt{2}}$, which implies that the singularity does not play a role if $\omega>\omega_S$. The fact that the singular arc occurs for $r_0=\omega\sqrt{2}$ and $s=0$ allows us to determine the position of the global optimum in the plane $(s,r_0)$ and $(\varphi_1,\varphi_2)$. These values are indicated in Fig.~\ref{fig5} and~\ref{fig6}. We observe the smooth continuity of the transition between the singular and the regular regimes. Figure~\ref{fig7} shows the trajectory of each spin $\vec{M}_i$ for $\omega>\omega_S$ (regular), $\omega<0.38$ (singular) and $\omega=\omega_S$.
\subsection{Selective inversion}\label{sec4b}
\begin{figure}[h!]
\includegraphics[scale=0.61]{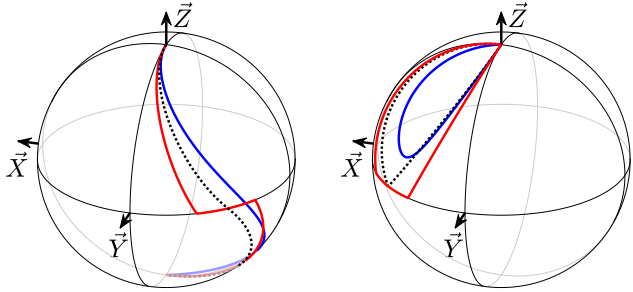}
\caption{(Color online) Same as Fig.~\ref{fig7} but in the case of the selective inversion process. The solid-blue (or solid dark gray), dotted-black and solid-red (or solid light gray) lines correspond to $\omega=0.8$, $\omega=\omega_S=1/\sqrt{2}$ and $\omega=0.5$, respectively. Dimensionless units are used.\label{fig8}}
\end{figure}

\begin{figure}[h!]
\centering
\includegraphics[scale=0.5]{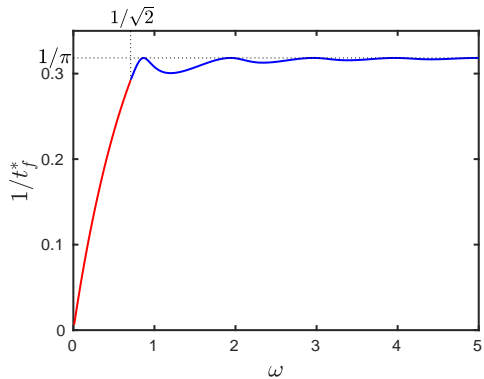}
\caption{(Color online) Inverse of the optimal time of the selective inversion process as a function of $\omega$. The red (light gray) and blue (dark gray) curves are associated to the singular and regular cases, respectively. The minima $t_f=\pi$ are obtained for $\omega=\tfrac{1}{2}\sqrt{4n^2-1}$ with $n\in\mathbb{N}$. Dimensionless units are used.\label{fig9}}
\end{figure}

For the selective inversion, the goal is to bring the spin $1$ to the south pole of the Bloch sphere, while keeping the position of the spin $2$ unchanged. We proceed as in Sec.~\ref{sec4a}. The figure of merit to minimize is:
\[J(t)=(1+z_1)^2+(1-z_2)^2.\]
In the singular case, we solve $z_1(t_f)=-1$ and $z_2(t_f)=1$ using Eq.~\eqref{blochsing}. We get:
\[
\begin{cases}
\Delta\alpha=\tfrac{\pi}{2},\\
T_s=\tfrac{\pi}{2\omega}-\tfrac{1}{\omega}\arctan\left(\tfrac{2\omega\sqrt{1-\omega^2}}{1-2\omega^2}\right),\\
t_f=\tfrac{2\arccos(-\omega^2)}{\sqrt{1+\omega^2}}+T_s.
\end{cases}
\]
Figure~\ref{fig8} shows the dynamics of the Bloch vectors for a regular solution, a singular one and for $\omega=\omega_S=1/\sqrt{2}$. Figure~\ref{fig9} displays the inverse of the final time as a function of $\omega$. For some specific offsets defined by $\omega=\tfrac{1}{2}\sqrt{4n^2-1}$ with $n\in\mathbb{N}$, we observe that the minimum time is $\pi$. This corresponds to a resonant inversion of the spin 1~\cite{wider}.
\section{Comparison with a direct numerical optimization}\label{sec5}
\begin{figure}[h!]
\centering
\includegraphics[scale=0.6]{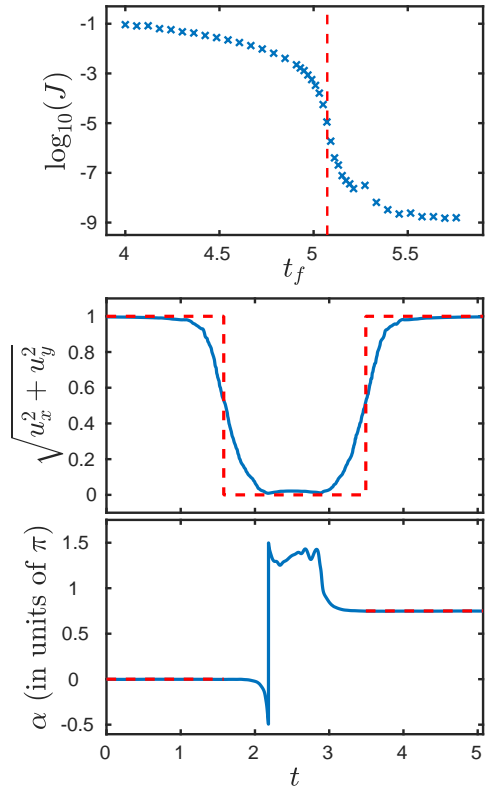}
\caption{(Color online) Numerical result obtained with GRAPE for a selective excitation process. The offset $\omega$ is set to $\omega=0.2$. Upper panel: Minimum figure of merit  $J(t_f)$ as a function of $t_f$. The red (light gray) dashed lines represent the optimal time obtained analytically with the singular solution [Eq~\eqref{solsingexc}]. Lower panels: Amplitude and phase of the control fields obtained numerically (blue or dark gray) and analytically (dashed-red or light gray). Dimensionless units are used.\label{fig10}}
\end{figure}

A specific attention must be paid to the singular solutions of the selective control of two spins. Since we do not have a proof of the optimality of these optimal solutions, we propose a numerical analysis with the GRAPE algorithm which is based on the PMP~\cite{grape}. This algorithm is able to deal with a large number of spins. The control of two spins can be treated very rapidly due to the low dimension of the control problem. A good estimation of the optimal trajectory can be achieved by running the algorithm with many different initializations. The goal of the optimization procedure is to minimize the figure of merit $J(t_f)=z_1^2(t_f)+(1-z_2(t_f))^2$, where $t_f$ is the final time, for an offest value $\omega=0.2$. We repeat the operation for different final times $t_f$ in a certain range in order to determine the minimum time $t_f$ for which $J(t_f)\simeq 0$. The result is shown in Fig.~\ref{fig10}. We can see that the figure of merit "falls" towards zero around $t=5.07$ which is the time corresponding to the singular analytical solution (Eq.~\eqref{solsingexc} with $\omega=0.2$). We observe in Fig.~\ref{fig10} that the numerical solution approaches the analytical one. The lack of precision is due to the used numerical optimization procedure in which the field is assumed to be a piecewise constant pulse. We refer the interested reader to Ref.~\cite{lapert:2010} for a complete discussion on this point.
\section{Discussion and prospective views}\label{sec6}
We have applied geometric optimal control techniques to the selective control in minimum time of two spins with different offsets. We have shown that the PMP leads to an illuminating interpretation of the optimal control problem in terms of a pseudo-particle whose dynamics are governed by an effective one-dimensional Hamiltonian. A geometric classification of regular and singular arcs is provided. We have finally described the time-optimal solution as a function of the offset parameters. Numerical results are presented both for the excitation and inversion processes. We have also recovered standard solutions used in NMR which are valid only for some specific offsets. Note that the same formalism and the same Pontryagin Hamiltonian can be used for other initial and final conditions. However, the fact that the two spins are initially on the north pole of the Bloch sphere greatly simplifies the analytical computations.

These results can be viewed as a first step toward the optimal selective control of spin systems. They also pave the way to other studies using the same approach, such as the selective excitation or inversion of three or more spins. In this case, the Pontryagin Hamiltonian may not be integrable and numerical shooting techniques~\cite{bonnardbook} have to used to find the optimal trajectory. The derivation of robust time optimal control fields for three and four spins (the initial and final states are the same for all the spins) has been made in~\cite{vandamme:2017}. State to state transfers are investigated in this work. It would be interesting to generalize this analysis to universal rotations, i.e. transfers which do not depend on the initial state of the system. Another interesting study would be to combine selectivity for some spins and robustness for the others. These two aspects will be addressed in a forthcoming paper~\cite{ansel:2018}.

\appendix
\section{Derivation of the regular arcs\label{appA}}
This paragraph is aimed at showing how to derive analytically the regular arcs which are solutions of Eq.~\eqref{eqkepler} and~\eqref{eqr2t}.
\subsection{Analytical expression\label{appA1}}
We define $u(t)=1/r(t)$. The integrals~\eqref{eqkepler} and~\eqref{eqr2t} become:
\begin{equation}
\omega st=\pm\int_{u_0}^{u}\frac{du'}{u'\sqrt{P_4(u')}},\quad \alpha=\pm\int_{u_0}^{u}\frac{u'du'}{\sqrt{P_4(u')}},\label{eqIntalphar}
\end{equation}
where:
\[
\small
\begin{aligned}
& P(u')=-u'^4+Au'^2+Bu-C,\\
& A=\tfrac{2\omega^2-r_0^2}{\omega^2s^2},\;
B=\tfrac{2r_0}{\omega^2s^2},\;
C=\tfrac{1+\omega^2}{\omega^2s^2}.
\end{aligned}
\]
We denote by $\beta_1$, $\beta_2$, $\gamma_1$, $\gamma_2$ the four roots of $P(u')$. $\beta_1$ and $\beta_2$ are real and ordered such that $\beta_1<u(t)<\beta_2$. $\gamma_1$ and $\gamma_2$ can be real or complex. These roots can be computed numerically or analytically by solving a four-order polynomial. In both cases, the solution is the sum of two elliptic integrals and a simple function. We set:
\begin{equation}
\begin{aligned}
&\omega s t=A_0I_0+A_1I_1+A_2I_2,\\ &\alpha=B_0J_0+B_1J_1+B_2J_2.
\end{aligned}
\end{equation}
The solution is given in Tab.~\ref{table1}. The elliptic integrals are defined by:
\[
\begin{aligned}
&\ellF(u,m)=\int_0^{u}\tfrac{d\theta}{\sqrt{1-m\sin^2\theta}}\\
&\Pi(n,u|m)=\int_0^u\tfrac{d\theta}{(1-n\sin^2\theta)\sqrt{1-m\sin^2\theta}},
\end{aligned}
\]
where the modulus $m$ belongs to $[0,1]$.
\begin{table*}\label{table1}
\footnotesize
\begin{tabular}{|c||c|}
\hline
If $(\gamma_1,\gamma_2)\in\mathbb{R}^2$ & If $(\gamma_1,\gamma_2)\in\mathbb{C}^2$\\
\hline\hline
 & \\
\underline{Solution} & \underline{Solution}\\
$
\begin{aligned}
&\begin{cases}
A_0=-\tfrac{4(1+\lambda_1)(\lambda_2-\lambda_1)}{(\beta_1+\beta_2)^2(1-\lambda_1)^2(1-n_a)\sqrt{k\lambda_2}} \\ A_1=\tfrac{2\sqrt{1+\lambda_1}(1+\lambda_2)^{3/2}}{(\beta_1+\beta_2)^2(\lambda_2-1)\sqrt{\lambda_2}}\\
A_2=-\tfrac{4\sqrt{(1+\lambda_1)(1+\lambda_2)}(\lambda_2-\lambda_1)}{(\beta_1+\beta_2)^2(1-\lambda_1)(\lambda_2-1)\sqrt{\lambda_2}}\\
I_0=\arctan(\sqrt{k}x)-\arctan(\sqrt{k}x_0)\\
I_1=\ellF\left(\sigma,\tfrac{\lambda_1}{\lambda_2}\right)-\ellF\left(\sigma_0,\tfrac{\lambda_1}{\lambda_2}\right)\\
I_2=\Pi\left(n_a;\sigma,\tfrac{\lambda_1}{\lambda_2}\right)-\Pi\left(n_a;\sigma_0,\tfrac{\lambda_1}{\lambda_2}\right)
\end{cases}\\
&\begin{cases}
B_0=-1\\
B_1=\tfrac{1-\lambda_2}{2\sqrt{\lambda_2}}\sqrt{\tfrac{1+\lambda_1}{1+\lambda_2}}\\
B_2= \tfrac{\lambda_2-\lambda_1}{\sqrt{\lambda_2(1+\lambda_2)(1+\lambda_1)}}\\
J_0=\arctan(\sqrt{\lambda_2}x)-\arctan(\sqrt{\lambda_2}x_0)\\
J_1=\ellF\left(\sigma,\tfrac{\lambda_1}{\lambda_2}\right)-\ellF\left(\sigma_0,\tfrac{\lambda_1}{\lambda_2}\right)\\
J_2=\Pi\left(n_b;\sigma,\tfrac{\lambda_1}{\lambda_2}\right)-\Pi\left(n_b;\sigma_0,\tfrac{\lambda_1}{\lambda_2}\right)
\end{cases}
\end{aligned}$ & $
\begin{aligned}
&\begin{cases}
A_0=-\tfrac{4(1+\lambda_1)(\lambda_1-\lambda_2)}{(\beta_1+\beta_2)^2(1-\lambda_1)^2(1-n_a)\sqrt{-k\lambda_2}} \\
A_1=-\tfrac{2\sqrt{1+\lambda_1}(-1-\lambda_2)^{3/2}}{(\beta_1+\beta_2)^2(1-\lambda_2)\sqrt{\lambda_1-\lambda_2}}\\
A_2=\tfrac{4\sqrt{(1+\lambda_1)(\lambda_1-\lambda_2)(-1-\lambda_2)}}{(\beta_1+\beta_2)^2(1-\lambda_1)(1-\lambda_2)(1-n)}\\
I_0=\arctan(\sqrt{k}x)-\arctan(\sqrt{k}x_0)\\
I_1=\ellF\left(\sigma,\tfrac{\lambda_1}{\lambda_1-\lambda_2}\right)-\ellF\left(\sigma_0,\tfrac{\lambda_1}{\lambda_1-\lambda_2}\right)\\
I_2=\Pi\left(\tfrac{n_a}{n_a-1};\sigma,\tfrac{\lambda_1}{\lambda_1-\lambda_2}\right)\\
-\Pi\left(\tfrac{n_a}{n_a-1};\sigma_0,\tfrac{\lambda_1}{\lambda_1-\lambda_2}\right)
\end{cases}\\
&\begin{cases}
B_0=1\\
B_1=\tfrac{(\lambda_2-1)\sqrt{1+\lambda_1}}{2\sqrt{(\lambda_2-\lambda_1)(1+\lambda_2)}}\\
B_2= \tfrac{-\lambda_2(1+\lambda_1)}{\sqrt{(\lambda_2-\lambda_1)(1+\lambda_2)(1+\lambda_1)}}\\
J_0=\arctan(\sqrt{-\lambda_2}x)-\arctan(\sqrt{-\lambda_2}x_0)\\
J_1=\ellF\left(\sigma,\tfrac{\lambda_1}{\lambda_2}\right)-\ellF\left(\sigma_0,\tfrac{\lambda_1}{\lambda_2}\right)\\
J_2=\Pi\left(n_b;\sigma,\tfrac{\lambda_1}{\lambda_2}\right)-\Pi\left(n_b;\sigma_0,\tfrac{\lambda_1}{\lambda_2}\right)
\end{cases}
\\
\end{aligned}$\\ & \\
\underline{Variables and parameters} & \underline{Variables and parameters} \\
$\begin{cases}
\sigma=\arcsin z\\
x=\textstyle\sqrt{\tfrac{\lambda_1(1-z^2)}{\lambda_2-\lambda_1z^2}},\\
z=\sqrt{\tfrac{\lambda_2(1+\lambda_1)}{\lambda_1(1+\lambda_2)}}\left(\tfrac{u-\tfrac{\beta_1+\beta_2}{2}\tfrac{1-\lambda_1}{1+\lambda_1}}{u-\tfrac{\beta_1+\beta_2}{2}\frac{1-\lambda_2}{1+\lambda_2}}\right)
\end{cases}
\begin{cases}
n_a=\tfrac{\lambda_1(1-\lambda_2)^2(1+\lambda_1)}{\lambda_2(1-\lambda_1)^2(1+\lambda_2)}\\
n_b=\tfrac{\lambda_1(1+\lambda_2)}{\lambda_2(1+\lambda_1)}\\
k=\tfrac{\frac{\lambda_2}{\lambda_1}n_a-1}{1-n_a}.
\end{cases}$ &
$\begin{cases}
\sigma=\arccos z,~n_a=\tfrac{\lambda_1(1-\lambda_2)^2(1+\lambda_1)}{\lambda_2(1-\lambda_1)^2(1+\lambda_2)}\\
x=\textstyle\sqrt{\tfrac{\lambda_1(1-z^2)}{-\lambda_2+\lambda_1z^2}},n_b=\tfrac{\lambda_1(1+\lambda_2)}{\lambda_2(1+\lambda_1)}\\
z=\sqrt{\tfrac{\lambda_2(1+\lambda_1)}{\lambda_1(1+\lambda_2)}}\left(\tfrac{u-\tfrac{\beta_1+\beta_2}{2}\tfrac{1-\lambda_1}{1+\lambda_1}}{u-\tfrac{\beta_1+\beta_2}{2}\frac{1-\lambda_2}{1+\lambda_2}}\right)\\
k=\tfrac{1-\frac{\lambda_2}{\lambda_1}n_a}{1-n_a}.
\end{cases}$
\\
\hline
\end{tabular}
\caption{Explicit expression of integrals~\eqref{eqIntalphar}.}
\end{table*}

The parameters $\lambda_1$ and $\lambda_2$ are given by:
\begin{equation}
\begin{cases}
\lambda_1=\tfrac{(\beta_1+\beta_2)^2+2\left(\beta_1\beta_2+\tfrac{C}{\beta_1\beta_2}\right)-2\sqrt{\delta}}{(\beta_1+\beta_2)^2-4\tfrac{C}{\beta_1\beta_2}}\\
\lambda_2=\tfrac{(\beta_1+\beta_2)^2+2\left(\beta_1\beta_2+\tfrac{C}{\beta_1\beta_2}\right)+2\sqrt{\delta}}{(\beta_1+\beta_2)^2-4\tfrac{C}{\beta_1\beta_2}}
\end{cases}\label{eqlambda12}
\end{equation}
with $\delta=(\beta_1\beta_2-C/(\beta_1\beta_2))^2+2(\beta_1+\beta_2)^2(\beta_1\beta_2+C/(\beta_1\beta_2))$. In the problem under study, we have $\lambda_2>1$ in the real case and $\lambda_2<-1$ in the complex case, while $\lambda_1\in[0,1]$ in both cases.
\subsection{Main steps of the demonstration\label{appA2}}
We give in this section the main steps to obtain the results given in Tab.~\ref{table1}. We focus on the case $(\gamma_1,\gamma_2)\in \mathbb{R}^2$ for sake of simplicity. The other case can be derived along the same lines. The method is described in Ref.~\cite{abramovitz,lawdenbook}. First, we express the four-order polynomial $P$ as the product of two polynomials of degree $2$. We set $P(u')=Q_1(u')Q_2(u')$ with:
\[
\begin{cases}
Q_1=-u^2+u(\beta_1+\beta_2)-\beta_1\beta_2\\
Q_2=u^2+u(\beta_1+\beta_2)+\tfrac{C}{\beta_1\beta_2},
\end{cases}
\]
where we have used the relations $\gamma_1\gamma_2=C/(\beta_1\beta_2)$ and $\beta_1+\beta_2=-(\gamma_1+\gamma_2)$. Then we express $Q_1$ and $Q_2$ as a sum of perfect squares. We compute the discriminant of $Q_1-\lambda Q_2$ and we determine the values of $\lambda$ which nullify this discriminant. We find $\lambda_1$ and $\lambda_2$ of Eq.~\eqref{eqlambda12}. We thus get an expression for $Q_1-\lambda_1 Q_2$ and $Q_1-\lambda_2 Q_2$. We arrive at:
\[
\begin{cases}
\begin{split}
Q_1=&-\tfrac{\lambda_2(1+\lambda_1)}{\lambda_2-\lambda_1}\left[u-\tfrac{\beta_1+\beta_2}{2}\tfrac{1-\lambda_1}{1+\lambda_1}\right]^2\\
 &+\tfrac{\lambda_1(1+\lambda_2)}{\lambda_2-\lambda_1}\left[u-\tfrac{\beta_1+\beta_2}{2}\tfrac{1-\lambda_2}{1+\lambda_2}\right]^2
 \end{split}\\
 \begin{split}
Q_2=&-\tfrac{1+\lambda_1}{\lambda_2-\lambda_1}\left[u-\tfrac{\beta_1+\beta_2}{2}\tfrac{1-\lambda_1}{1+\lambda_1}\right]^2\\
&+\tfrac{1+\lambda_2}{\lambda_2-\lambda_1}\left[u-\tfrac{\beta_1+\beta_2}{2}\tfrac{1-\lambda_2}{1+\lambda_2}\right]^2.
\end{split}
\end{cases}
\]
Finally, the change of variables $$
z=\sqrt{\tfrac{\lambda_2(1+\lambda_1)}{\lambda_1(1+\lambda_2)}}\left(\tfrac{u-\tfrac{\beta_1+\beta_2}{2}\tfrac{1-\lambda_1}{1+\lambda_1}}{u-\tfrac{\beta_1+\beta_2}{2}\frac{1-\lambda_2}{1+\lambda_2}}\right)
$$
allows us to write $Q_1$ and $Q_2$ in a simpler form. This change of variables is made in the integrals of Eq.~\eqref{eqIntalphar}. For example, the first integral becomes:
\[\small
\omega st=2\tfrac{(1+\lambda_1)(1+\lambda_2)\sqrt{\lambda_1}}{(\beta_1+\beta_2)^2(1-\lambda_1)\lambda_2}\int_{z_0}^z\tfrac{z-\sqrt{\tfrac{\lambda_2(1+\lambda_1)}{\lambda_1(1+\lambda_2)}}}{\left(1+z\sqrt{n_a}\right)\sqrt{\left(1-\tfrac{\lambda_1}{\lambda_2}z^2\right)(1-z^2)}}.
\]
Multiplying the nominator and denominator by $1-z\sqrt{n_a}$, we can express this integral as a linear combination of three integrals, two of them are elliptic and the other one can be expressed with simple functions. More precisely, we obtain the three following integrals:
\[
\begin{aligned}
& F_0=\int_{z_0}^z\tfrac{zdz}{\left(1-n_az^2\right)\sqrt{\left(1-\tfrac{\lambda_1}{\lambda_2}z^2\right)(1-z^2)}},\\
& F_1=\int_{z_0}^z\tfrac{dz}{\sqrt{\left(1-\tfrac{\lambda_1}{\lambda_2}z^2\right)(1-z^2)}},\\
& F_2=\int_{z_0}^z\tfrac{dz}{\left(1-n_az^2\right)\sqrt{\left(1-\tfrac{\lambda_1}{\lambda_2}z^2\right)(1-z^2)}}.
\end{aligned}
\]
Making the change of variables $x=\textstyle\sqrt{\tfrac{\lambda_1(1-z^2)}{\lambda_2-\lambda_1z^2}}$ in $F_0$ and $z=\sin\sigma$ in $F_1$ and $F_2$, we get the result of Tab.~\ref{table1}.
\subsection{Case $s=0$\label{appA3}}
In this case, Eq.~\eqref{eqkepler} and~\eqref{eqr2t} can be expressed in terms of simple analytical functions. If $E>0$, the pseudo-particle crosses the singular point $r=0$ at some specific times, leading to a jump of $\pi$ for the control phase. The solution is given in Tab.~\ref{table2}.
\begin{table}
\begin{center}
\footnotesize
\begin{tabular}{|c|}
\hline
If $E<0$ ($r_0>\omega\sqrt{2}$)\\
\hline
$\begin{cases}
r(t)=\textstyle\left.\left.\tfrac{1}{1+\omega^2}\right(r_0-\omega\sqrt{2(1+\omega^2)-r_0^2}\cos u\right) \\
\alpha(t)=0
\end{cases}$\\
\hline
$\begin{aligned}
&u=\textstyle\sqrt{1+\omega^2}t+\rho,\\
&\rho=\textstyle-\arccos\left(\tfrac{-\omega r_0}{\sqrt{2(1+\omega^2)-r_0^2}}\right)
\end{aligned}$\\
\hline\hline
If $E>0$ ($r_0<\omega\sqrt{2}$)\\
For $t\in[t_{n-1},t_{n}]$ with $n\in\mathbb{N}\backslash 0$\\
\hline
$\begin{cases}
r(t)=\textstyle\left.\left.\tfrac{1}{1+\omega^2}\right(r_0-\omega\sqrt{2(1+\omega^2)-r_0^2}\cos u\right) \\
\alpha(t)=(n-1)\pi
\end{cases}$\\
\hline
$\begin{cases}
u(t)=\textstyle\sqrt{1+\omega^2}t+\rho_{n-1}\\
\rho_n=\textstyle 2n\arccos\left(\tfrac{r_0}{\omega\sqrt{2(1+\omega^2)-r_0^2}}\right)+\rho_0\\
t_n=\textstyle \tfrac{1}{\sqrt{1+\omega^2}}\left(2(n-1)\pi-\arccos\left(\tfrac{r_0}{\omega\sqrt{2(1+\omega^2)-r_0^2}}\right)-\rho_{n-1}\right),\\
\rho_0=\textstyle -\arccos\left(\tfrac{-\omega r_0}{\sqrt{2(1+\omega^2)-r_0^2}}\right),\\
t_0=0.
\end{cases}$\\
\hline
\end{tabular}
\end{center}
\caption{Solution of Eq.~\eqref{eqkepler} and~\eqref{eqr2t} for $s=0$.\label{table2}}
\end{table}
\section{Derivation of the singular control fields\label{appB}}
In this section, we compute the control fields $u_x$ and $u_y$ in the singular case, which occurs when $r(t)=0$ over a finite time. In the coordinates $\vec{\ell}=\vec{L}_1+\vec{L}_2$ and $\vec{m}=\vec{L}_1-\vec{L}_2$, the pseudo-hamiltonian can be expressed as:
$$
H_p=u_x\ell_x+u_y\ell_y-\omega m_z.
$$
The Hamilton's equations lead to the following relations:
\begin{equation*}
\begin{cases}
\dot{\ell}_x=-\omega m_y-u_y\ell_z \\
\dot{\ell}_y=\omega m_x+u_x\ell_z \\
\dot{\ell}_z=u_y\ell_x-u_x\ell_y,
\end{cases}
\end{equation*}
and
\begin{equation*}
\begin{cases}
\dot{m}_x=-\omega \ell_y-u_y m_z \\
\dot{m}_y=\omega \ell_x+u_x m_z \\
\dot{m}_z=u_y m_x-u_x m_y.
\end{cases}
\end{equation*}
The singular set is defined by $\ell_x(t)=\ell_y(t)=0$ ($r=\sqrt{\ell_x^2+\ell_y^2}=0$). We also know that $\ell_z$ is a constant of motion which is zero for the initial point of the dynamics. We deduce that $m_x=m_y=0$ on the singular set. We have therefore two cases to consider:
\begin{itemize}
\item $m_z=0$ and thus $H_p=0$. It corresponds to the exceptional case which does not appear in this case.
\item $u_x(t)=u_y(t)=0$ and $m_z$ constant, which can be different from zero.
\end{itemize}
As explained in Sec.~\ref{sec3b}, the singular case exists only if $s=0$. Thus, a regular-singular solution is a succession of constant pulses of maximum and zero amplitudes.
\section{Dynamics of the Bloch vector\label{appC}}
\subsection{Regular case\label{appC1}}
We can show that the Bloch vector $\vec{M}_i$, the angular momentum $\vec{L}_i$ and the adjoint state $\vec{p}_i$ of the spin $i$ form an orthogonal basis which can be described by three Euler angles called $\theta_i$, $\phi_i$ and $\psi_i$~\cite{goldsteinbook}. We define $\theta_i$ and $\phi_i$ such that:
\begin{equation}
\begin{cases}
L_{x_i}=L_i\sin\theta_i\cos\phi_i,\\
L_{y_i}=L_i\sin\theta_i\sin\phi_i,\\
L_{z_i}=L_i\cos\theta_i,
\end{cases}\label{eqLiEA}
\end{equation}
where $L_i=|\vec{L}_i|$. The third Euler angle describes the motion of $\vec{M}_i$ and $\vec{p}_i$. The bloch vector $\vec{M}_i$ is defined as:
\begin{equation}
\begin{cases}
x_i=-\cos\phi_i\sin\psi_i\cos\theta_i-\sin\phi_i\cos\psi_i,\\
y_i=-\sin\phi_i\sin\psi_i\cos\theta_i+\cos\phi_i\cos\psi_i,\\
z_i=\sin\psi_i\sin\theta_i.
\end{cases}\label{eqMiEA}
\end{equation}
Knowing the dynamics of the Bloch vector [Eq.~\eqref{eqBloch}] and of the angular momenta [Eq.~\eqref{eqmomcin}], we can find the dynamics of each Euler angle. In particular, the dynamics of $\psi_i$ can be derived from:
\begin{equation}
\dot{\psi}_i=-\tfrac{u_x\cos\phi_i+u_y\sin\phi_i}{\sin\theta_i},
\label{eqdpsi2t}
\end{equation}
where $u_x$ and $u_y$ are given by Eq.~\eqref{optfields}. The angles $\theta_i$ and $\phi_i$ can be expressed by inverting Eq.~\eqref{eqLiEA}, since $\theta_i\in ]0,\pi[$. We arrive at:
\[\begin{cases}
\sin\theta_i=\tfrac{\sqrt{L_{x_i}^2+L_{y_i}^2}}{L_i} \\
\cos\theta_i=\tfrac{L_{z_i}}{L_i}\\
\sin\phi_i=\tfrac{L_{y_i}}{\sqrt{L_{x_i}^2+L_{y_i}^2}}\\
\cos\phi_i=\tfrac{L_{x_i}}{\sqrt{L_{x_i}^2+L_{y_i}^2}}.
\end{cases}\]
The change of coordinates to $\vec{\ell}$, $\vec{m}$ and $\vec{r}$ leads to Eq.~\eqref{blochreg} and~\eqref{eqdpsidtreg}.
\subsection{Singular case\label{appC2}}
We recall that, in this case, the pulse is the concatenation of a regular pulse of phase $\alpha=0$, a zero-amplitude pulse and a second regular arc of phase $\Delta\alpha$. The two regular arcs are of the same duration $T_r$ (see Fig.~\ref{fig2}), and the singular arc lasts during the time $T_s$.

A constant control pulse of duration $\Delta t$ leads to a rotation of angle $\sqrt{u_x^2+u_y^2+\omega_i^2}\Delta t$ about $\vec{n}={}^t(u_x,u_y,\omega_i)$. Starting from $t=t_a$ and $\vec{M}_a=(x_a,y_a,z_a)$ until the time $t_b$ with $t_b-t_a=\Delta t$, the explicit solution of the Bloch equation can be expressed as:
\begin{equation}
\small
\begin{cases}
\begin{aligned}
x_b=&uu_{x}[x_au_{x}+y_au_{y}]+z_a[vu_{x}-wu_{y}]\\
&+\tfrac{\omega_i}{\Omega}y_a\sin(\Omega\Delta t)+x_a\cos(\Omega\Delta t)
\end{aligned}
\\
\begin{aligned}
y_b=&uu_{y_n}[x_au_{x}+y_au_{y}]+z_a[vu_{y}+wu_{x}]\\
&-\tfrac{\omega_i}{\Omega}x_a\sin(\Omega\Delta t)+y_a\cos(\Omega\Delta t)
\end{aligned}\\
z_b=u_{y}[w x_a+vy_a]+u_{x}[vx_a-wy_a]+z_a(1-u).
\end{cases}\label{eqsolcstpulse}
\end{equation}
with:
\[\begin{array}{ll}
\Omega=\sqrt{\omega_i^2+u_x^2+u_y^2}, &u=\tfrac{1}{\Omega^2}\big(1-\cos(\Omega\Delta t)\big),\\
v=\tfrac{\omega_i}{\Omega^2}\big(1-\cos(\Omega\Delta t)\big),& w=\tfrac{1}{\Omega}\sin(\Omega\Delta t).
\end{array}
\]
Thus, the solution of the Bloch equation driven by the regular-singular control pulse is given by the concatenation of the solution associated to $u_x=1$ and $u_y=0$ from $t=0$ to $T_r$, $u_x=u_y=0$ from $T_r$ to $T_r+T_s$, $u_x=\cos(\Delta\alpha)$ and $u_y=\sin(\Delta\alpha)$ from $T_r+T_s$ to $2T_r+T_s$. Substituting $T_r$ by $\arccos(-\omega^2)/\sqrt{1+\omega^2}$ and $\gamma=\arctan[2\omega\sqrt{1-\omega^2}/(1-2\omega^2)]$, we get the solution of Eq.~\eqref{blochsing}.\\

\noindent\textbf{ACKNOWLEDGMENT}\\
S.J. Glaser acknowledges support from the DFG (Gl 203/7-2). D. Sugny and S. J. Glaser acknowledge support from the ANR-DFG research program Explosys (ANR-14-CE35-0013-01). D. Sugny acknowledges support from the PICS program, from the ANR-DFG research program COQS (ANR-15-CE30-0023-01) and from the ANR research
program QUACO (ANR-17-CE40-0007-01). The work of D. Sugny has been done with the support of the Technische Universit\"at M\"unchen – Institute for Advanced Study, funded by the German Excellence Initiative and the European Union Seventh Framework Programme under grant agreement 291763. This project has received funding from the European Union's Horizon 2020 research and innovation programme under the Marie-Sklodowska-Curie grant
agreement No 765267.

\end{document}